\documentclass[aps,prl,twocolumn,showpacs]{revtex4}

\usepackage{amsmath}
\usepackage{amssymb}
\usepackage{graphicx}% Include figure files
\usepackage{dcolumn}% Align table columns on decimal point
\usepackage{bm}% bold math

\usepackage{epsfig}
\usepackage{graphics}

\newcommand{\eps}{\varepsilon}
\newcommand{\la}{\lambda}
\newcommand{\K}{{\mathrm{K}} }
\newcommand{\E}{{\mathrm{E}} }
\newcommand{\cn}{{\mathrm{cn}} }
\newcommand{\dn}{{\mathrm{dn}} }
\newcommand{\sn}{{\mathrm{sn}} }

\begin{document}
%\twocolumn[\hsize\textwidth\columnwidth\hsize
%\csname@twocolumnfalse%
%\endcsname
%\draft

\title{
Adiabatic dynamics of periodic waves in Bose-Einstein condensate
with time dependent atomic scattering length}
\author{F. Kh. Abdullaev$^1$}
%\email{fatkh@physic.uzsci.net}
\author{A.M. Kamchatnov$^{2,3}$}
%\email{kamch@isan.troitsk.ru}
\author{V. V. Konotop$^3$}
%\email{konotop@cii.fc.ul.pt}
\author{V. A. Brazhnyi$^3$}
\affiliation{
$^1$Physical-Technical Institute of the Uzbek Academy of
Sciences,2-b, Mavlyanov str., 700084, Tashkent, Uzbekistan \\
$^2$Institute of Spectroscopy, Russian Academy of Sciences, Troitsk, Moscow Region,
142190, Russia \\
$^3$Centro de F\'{\i}sica Te\'orica e Computacional, Universidade de Lisboa,  Av.~Prof.~Gama Pinto 2, Lisbon 1649-003,
Portugal}

\date{\today}

\begin{abstract}
Evolution of periodic  matter waves in one-dimensional
Bose-Einstein condensates with time dependent scattering length is
described. It is shown that variation of the effective
nonlinearity is a powerful tool for controlled generation of
bright and dark solitons starting with periodic waves.
\end{abstract}


\maketitle

Observation of Bose-Einstein condensate (BEC) in gases of weakly
interacting alkali metals have stimulated intensive studies of the
nonlinear matter waves. A new area of physics---nonlinear matter
waves and nonlinear atomic optics---was originated. Generation and
dynamics of solitary wave pulses in BEC's is one of the most important
related problems. Experimental observations of dark~\cite{dark,Den} and
bright~\cite{Khayk,Hulet} solitons have recently been reported.
Theoretically, several methods of creating solitary waves have been
proposed. First of all, it is a modulational instability~\cite{Konotop},
which is a universal phenomenon of the nonlinear physics (especially
intensively explored in nonlinear optics, see \cite{AbdRev}). This method,
however, cannot predict exactly the parameters of generated solitons. Another
method, which is controllable in the above sense, is the so-called
phase engineering~\cite{Den}, which consists of imposing an initial
phase on a BEC and is appropriate for generating dark solitons. The phase
imprinting, however, affects the whole background condensate which acquires
nonzero initial velocity and starts to oscillate in a trap potential.
The problem becomes even more complicated when one is interested in
generating trains (or lattices) of solitons in BEC's.

In the present Letter we show that a powerful tool for generating and
managing  matter soliton trains can be provided by variation of the effective
nonlinearity, which in practical terms can be achieved by means of
variation of the $s$-wave scattering length $a_{s}$ due to the Feshbach
resonance \cite{Moer}:
\begin{equation}\label{SL}
a_{s}(t) = a(1 + \Delta/(B_{0} - B(t)).
\end{equation}
Here $a$ is the asymptotic value of the scattering length far from resonance,
$B(t)$ is the time-dependent external magnetic field, $\Delta$ is the width
of the resonance and $B_{0}$ is the resonant value of the magnetic field.
Feshbach resonances have been observed in Na at 853 and 907 G \cite{Inouye},
in $^7$Li at 725 G \cite{Hulet}, and in $^{85}$Rb at 164 G with $\Delta = 11$G \cite{Court}. Also, rapid variation in time of
$a_{s}$ has been recently used for generation of bright solitons in
BEC \cite{Khayk,Hulet}. Here we want to indicate that in quasi-one-dimensional
geometry an initially weak modulation of the condensate can be amplified by
means of proper variation of the scattering length. As a result, the
condensate evolves into either a sequence of bright solitons for $a_s<0$,
or ``domains'' separated by dark solitons for $a_s>0$.
In the case of bright solitons the attractive forces between atoms exactly compensate the wave-packet dispersion in the longitudinal direction, so that the confining trap potential in this direction becomes unnecessary. Then the motion of bright solitons in
the longitudinal direction can be controlled by means of application of
external forces. Actually, oscillations of bright solitons in the trap
observed in \cite{Khayk,Hulet} give a simple example of such controllable
motion. Thus, quasi-one-dimensional bright BEC solitons behave as separate
entities and their investigation seems to be a quite promising field of
research.
Dark solitons may be considered as moving ``domain walls''
which separate regions of a condensate with different values of the
order parameter. Investigation of dark solitons is also useful for
understanding the properties of BEC.
In general, the problem of the controllable soliton
generation is important for a number of BEC applications, like
atomic interferometry~\cite{Wright}, and different kinds
of the quantum phase transitions \cite{Ueda}, as well as in the context
of the nonlinear physics, including nonlinear optics and hydrodynamics.

Our approach is based on the well established concept that the BEC dynamics
at low enough temperature is well described by the three-dimensional (3D)
Gross-Pitaevskii (GP) equation.  In some physically important cases it admits a
self-consistent reduction to the 1D nonlinear Schr\"{o}dinger
(NLS) equation
\begin{equation}\label{NLS}
iu_{t} + u_{xx} - \frac 12 \nu^2x^2 u - 2\sigma g|u|^{2}u= 0.
\end{equation}
In particular this is the case of  a cigar-shaped BEC of low
density when
$\epsilon=\frac{{\cal N}|a|}{a_\bot}\ll 1$ and $\frac{a_\bot^2}{a_0^2}=\frac{\epsilon^2\nu}{\sqrt{2}}$, where ${\cal N}$ is a total number of atoms, $a_\bot=(\frac{\hbar}{m_a\omega_\bot})^{1/2}$ and $a_0=(\frac{\hbar}{m_a\omega_0})^{1/2}$ are linear oscillator lengths in the transverse and in cigar-axis directions, respectively (in the small amplitude limit they are of order of effective sizes of the condensate), $\omega_\bot$ and $\omega_0$ being
respective harmonic oscillator frequencies, $\nu\lesssim 1$ is a positive parameter , and $m_a$ is the atomic mass. In (\ref{NLS}) time $t$ and coordinate $x$ are measured in   units $2/(\epsilon^2\omega_\bot)$ and $a_\bot/\epsilon$ respectively. The order parameter in the leading order is searched in the form $\psi({\bf r}, t)=\frac{\epsilon}{\sqrt{2\pi |a|}a_\bot}\exp\left(-i\omega_\bot t-\frac{y^2+ z^2}{2a_\bot^2}\right)u\left(\frac{\epsilon x}{a_\bot}, \frac{\epsilon^2\omega_\bot t}{2}\right)$, where $\sigma={\rm sign}(a_s)$, and $g(t)\equiv a_s(t)/a_s(0)$. It will be assumed that $a_s(t)$ does not change its sign and thus $g(t)$ is a positive-valued function. We notice that smallness of the density rules out a possibility of collapse phenomenon (if $a<0$). 

We start with analytical estimates supposing that the initial wave function
$u(x,0)$ is modulated along the cigar axis with the wavelength $L$ of
modulation much less than the longitudinal dimension of the condensate,
i.e. of the $l$: $L\ll l$. Therefore, at this stage we neglect the smooth
trap potential and impose cyclic boundary conditions.  Then the initial
wave function can be approximated well
enough by exact periodic solutions of Eq.~(\ref{NLS}) at $\nu=0$. For example,
if at $t=0$ we take into account only one space harmonic  of the initial
wave function,  $u(x,0)=u_0+u_1\cos(x/L)$, then this distribution can
be approximated by  well-known elliptic function solutions of
Eq.~(\ref{NLS}) with a small parameter $m$ (see below). We are interested
in evolution of such solutions due to slow change of $g(t)$ with time.
At the same time we suppose that the total time of adiabatically slow
change of $g(t)$ is much less than the period $\sim2\pi/\nu$ of soliton
oscillations in the trap potential, so that we can neglect the influence
of the trap potential on the motion of solitons during the formation of
soliton trains and put $\nu=0$ in Eq.~(\ref{NLS}). This means that we
shall consider analytically relatively small segments of the modulated wave
much greater than the wavelength $L$ and much smaller than the size $l$
of the whole condensate in the trap.
To solve the problem of the condensate evolution, we note that substitution
\begin{equation}\label{u-v}
  u(x,t)=v(x,t)/\sqrt{g(t)}
\end{equation}
transforms Eq.~(\ref{NLS}) with $\nu=0$ into
\begin{equation}\label{modNLS}
  iv_t+v_{xx}-2\sigma |v|^2v=i\eps v
\end{equation}
with
%\begin{equation}\label{eps}
$
  \eps(t)=g'(t)/2g(t).
$
%\end{equation}
Thus, for slowly varying $g(t)$ the right hand side of Eq.~(\ref{modNLS})
can be considered as a small perturbation: $|\varepsilon(t)|\ll 1$.
As it follows from Eq.~(\ref{u-v}), for the initial distribution one
has $v(x,0)=u(x,0)$. For our purposes it is enough to consider typical
particular solutions of the unperturbed NLS equation which are parameterized by two parameters $\la_{1,2}$~\cite{com}. Under influence of the perturbation, these parameters in the adiabatic approximation become slow functions of time $t$. Equations which govern their evolution can be derived by the following simple method.

First, the initial values of $\la_{1,2}$, as well as the coefficients in
Eq.~(\ref{modNLS}) are supposed to be independent of $x$, hence the wavelength
$L$ of the
nonlinear wave evolving according to Eq.~(\ref{modNLS}) is constant,
\begin{equation}\label{wl}
  {dL(\la_1(t),\la_2(t))}/{dt}=0.
\end{equation}
Second, we find that the variable
%\begin{equation}\label{N}
$ 
  N(\la_1(t),\la_2(t))=\int_0^L|v|^2dx
$
%\end{equation}
changes with time according to
\begin{equation}\label{Neq}
  {dN(\la_1(t),\la_2(t))}/{dt}=2\eps N(\la_1(t),\la_2(t)).
\end{equation}
Then, if the expressions for $L$ and $N$ in terms of $\la_{1,2}$ are
known, Eqs.~(\ref{wl}) and (\ref{Neq}) reduce to two equations linear
with respect to derivatives $d\la_{1,2}/dt$, which
yield the desired system of differential equations for $\la_{1,2}$.
The form of this system depends, of course, on the choice of the parameters
$\la_{1,2}$. It is well known from the theory of modulations of
nonlinear periodic waves that for completely integrable equations
(as the NLS equation) the most convenient choice is provided by the
so-called ``finite-gap integration method'' of obtaining periodic
solutions. Therefore we shall use the parametrization of the periodic
solutions of the NLS equation obtained by this method (see, e.g.
\cite{Kamch2000}), and consider three most typical cases.

{\em Case 1}: cn-{\em wave in a BEC with a negative scattering length}.
In the case of a BEC with negative scattering length, $\sigma=-1$,
there are two simple two-parametric periodic solutions of unperturbed
Eq.~(\ref{modNLS}). One of them has the form
\begin{equation}
\label{cn}
  v(x,t)=2\la_1e^{-4i(\la_1^2-\la_2^2)t}\cn[2\sqrt{\la_1^2+\la_2^2}\, x,m],
\end{equation}
where the parameter of elliptic function is given by
%\begin{equation}\label{m-cn}
$m={\la_1^2}/({\la_1^2+\la_2^2})$.
%\end{equation}
Then straightforward calculations give
\begin{equation}\label{L-cn}
  L=\frac{2\K(m)}{\sqrt{\la_1^2+\la_2^2}}, \quad
  N=8\sqrt{\la_1^2+\la_2^2}\,\E(m)-4\la_2^2 L
\end{equation}
where $\K(m)$ and $\E(m)$ are complete elliptic integrals of the first and
the second kind, respectively. Substitution of these expressions into
Eqs.~(\ref{wl}) and (\ref{Neq}) yields the system
\begin{equation}\label{cn-syst}
  \begin{array}{l}\displaystyle{
  \frac{d\la_1}{dz}=\frac{((\la_1^2+\la_2^2)\E(m)-\la_2^2\K(m))\E(m)\la_1}
  {\la_1^2\E^2(m)+\la_2^2(\K(m)-\E(m))^2},}\\
  \displaystyle{
  \frac{d\la_2}{dz}=\frac{(\la_2^2\K(m)-(\la_1^2+\la_2^2)\E(m))(\K(m)-\E(m))\la_2}
  {\la_1^2\E^2(m)+\la_2^2(\K(m)-\E(m))^2}}
  \end{array}
\end{equation}
where
\begin{equation}\label{z}
  z=z(t)=2\int_0^t\eps(t') dt'=\ln g(t), \qquad z(0)=0.
\end{equation}
If dependence of $\la_1$ and $\la_2$ on $z$ is found from (\ref{cn-syst}),
then Eq.~(\ref{u-v}) gives evolution of the periodic wave $u(x,t)$ with
slow change of the parameter $z$ connected with time $t$ by Eq.~(\ref{z}).
In particular, the density of particles in the condensate is given by
\begin{equation}\label{cn-dens}
|u|^2=4e^{-z}\la_1^2(z)\,\cn^2[2\sqrt{\la_1^2(z)+\la_2^2(z)}\, x,m],
\end{equation}
where transformation to the time variable should be performed with the
use of Eq.~(\ref{z}).

\begin{figure}[t]
%\vspace{0.2 true cm}
\centerline{\includegraphics[width=5cm,height=8cm,clip]{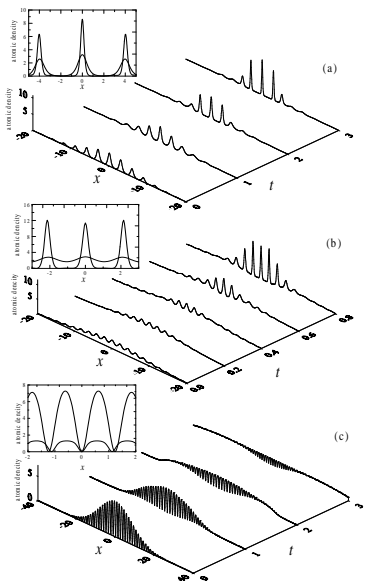}}
%\vspace{0.4 true cm}
\caption{Numerical solution of Eq.(\ref{NLS}) with $g(t)=e^{t/\tau}$.  
Initial conditions are chosen in the form $u(x,0)e^{-\nu x^2/2^{3/2}}$ where $u(x,0)$ is  given by 
(a) Eq. (\ref{cn}) with $\lambda_1(0)=1$, $\lambda_2(0)=.2$, $\tau\approx 2$
$\nu=.02$; 
(b) Eq. (\ref{dn}) with $\lambda_1(0)=1.5$, $\lambda_2(0)=.2$, $\tau\approx 1.5$
and $\nu=.01$;
(c) Eq. (\ref{sn}) with $\lambda_1(0)=3$, $\lambda_2(0)=.3$, $\tau\approx 1$
and $\nu=.01$.
In the boxes we show density distributions at initial (thin lines) and final (thick lines) moments of time.  
}
\label{figone}
\end{figure}

In Fig.~\ref{figone}a we present an example of
the evolution of the respective density distribution in the presence
of a harmonic trap potential where the parabolic parameter as well as experimentally feasible parameters are used. The figure shows
that in the case of a negative scattering length given by (\ref{SL})
increase of the magnetic field $B(t)$ within the region
$B(0)<B(t)<B_0$ results in compression of the atomic density and
formation of a lattice of matter solitons.

{\em Case 2}: dn-{\em wave in a BEC with a negative scattering length}.
Another simple solution of the NLS equation (\ref{modNLS}) with $\sigma=-1$
is given by
\begin{equation}
\label{dn}
 v(x,t)=(\la_1+\la_2)e^{-2i(\la_1^2+\la_2^2)t}\dn[(\la_1+\la_2)x,m],
\end{equation}
where
%\begin{equation}\label{m-dn}
$ m={4\la_1\la_2}/{(\la_1+\la_2)^2}$.
%\end{equation}
By analogy with (\ref{cn-syst}) we derive the following equations for $\la_1$
and $\la_2$:
\begin{equation}\label{dn-syst}
  \begin{array}{l}\displaystyle{
  \frac{d\la_1}{dz}=\frac{\la_1(\la_1+\la_2)\E(m)}
  {(\la_1-\la_2)\K(m)+(\la_1+\la_2)\E(m)},}\\
  \displaystyle{
  \frac{d\la_2}{dz}=-\frac{\la_2(\la_1+\la_2)\E(m)}
  {(\la_1-\la_2)\K(m)-(\la_1+\la_2)\E(m)},}
  \end{array}
\end{equation}
where it is supposed that $\la_1>\la_2$ and $z$ is defined by Eq.~(\ref{z}).
Now the density of particles is given by
\begin{equation}
\label{dn-dens}
|u|^2=e^{-z}(\la_1(z)+\la_2(z))^2\dn^2[(\la_1(z)+\la_2(z)) x,m].
\end{equation}
An example of the respective evolution in the presence of the potential is given in Fig.~\ref{figone}b. One again observes formation of a lattice of matter solitons
starting with a weakly modulated periodic wave.

{\em Case 3}: sn-{\em wave in a BEC with a positive scattering length}.
In the case of (\ref{modNLS}) with $\sigma=1$ there exists simple periodic
solution
\begin{equation}\label{sn}
  v(x,t)=(\la_1-\la_2)e^{2i(\la_1^2+\la_2^2)t}\sn[(\la_1+\la_2)x,m],
\end{equation}
where
$
%\begin{equation}\label{m-sn}
  m=\left(\frac{\la_1-\la_2}{\la_1+\la_2}\right)^2
%\end{equation}
$
and it is supposed that $\la_1>\la_2$. Now we obtain
\begin{equation}\label{sn-syst}
  \begin{array}{l}\displaystyle{
  \frac{d\la_1}{dz}=\frac{\la_1(\la_1+\la_2)(\K(m)-\E(m))}
  {2\la_1\K(m)-(\la_1+\la_2)\E(m)},}\\
  \displaystyle{
  \frac{d\la_2}{dz}=\frac{\la_2(\la_1+\la_2)(\K(m)-\E(m))}
  {2\la_2\K(m)-(\la_1+\la_2)\E(m)},}
  \end{array}
\end{equation}
with $z$ defined again by Eq.~(\ref{z}). The density of particles in the
condensate is given by
\begin{equation}\label{sn-dens}
|u|^2=e^{-z}(\la_1(z)-\la_2(z))^2\sn^2[(\la_1(z)+\la_2(z)) x,m].
\end{equation}
This case, but in the presence of external trap potential is illustrated
in Fig.~\ref{figone}c, where by means of increase of the
magnetic field a periodic wave is transformed into a lattice of
dark solitons.

In physical units, the cases depicted in Fig.~\ref{figone} correspond to  
(a) ${\cal N}= 1.4\cdot 10^4$ $^7Li$ atoms in a trap with $a_\bot\approx 7\,\mu$m, and $a_0\approx 230\,\mu$m, 
(b) ${\cal N}= 2\cdot 10^4$ $^7Li$ atoms in a trap with $a_\bot\approx 6\,\mu$m, and $a_0\approx 416\,\mu$m, and 
(c) ${\cal N}= 10^4$  $^{23}Na$ atoms in a trap with $a_\bot=3.4\,\mu$m, and $a_0=264\,\mu$m. 
In the last case, however one observes shifts of the soliton positions as well as decrease of a density of particles located about the potential minimum because of weak oscillations of the condensate in the trap potential (the expanding phase is depicted in the figure). Thus, although effective $\tau$ corresponding to physical time~\cite{Hulet} $t_0=40\,$ms  (used in all simulations) is not large characteristic amplitudes of solitons placed in the center of the trap potential (in the cases (a) and (b) match well the amplitude values following
from the adiabatic approximation developed for a homogeneous NLS and for large $\tau$, and in the case (c) one observes qualitative agreement.  No instabilities of periodic waves are observed during periods of soliton train formation.

The developed analytical approach can be generalized to the NLS equation
with linear damping, when the right hand side 
of Eq.~(\ref{NLS}) is equal to $-i\gamma u$, $\gamma$ being the
damping constant, the substitution (\ref{u-v}) yields again
Eq.~(\ref{modNLS}) but now with modified value of $\eps$:
$\eps\to\eps-\gamma$. Hence, the equations for $\la_1$ and $\la_2$
hold their form with $z(t)$ defined as $z(t)=\ln g(t)-2\gamma t$,
so that the only change in Eqs. (\ref{cn-dens}), (\ref{dn-dens}),
and (\ref{sn-dens}) is multiplication of their right hand sides by
$\exp(-2\gamma t)$. Another generalization corresponds to moving soliton 
trains what may be useful for treatment of BEC in ring traps.  The above method
remains actually the same, but the initial condition includes a
phase factor linearly depending on the space coordinate. Then the
results presented in Fig.~\ref{figone} can be viewed as plots of the respective
current as functions of time. In practice, solitons can be put also in
motion by imposing a proper external potential. One of applications
of such moving soliton trains could be a ``laser'' of matter solitons.

The consideration provided above implies that one starts the adiabatic deformation
with an initially periodic solution. A question arises about possibility of creation of such a state experimentally. A natural approach to solving this problem would be the use of an optical
trap \cite{opt_trap}. In such a trap it is possible to create a nonlinear periodic
distribution of a BEC \cite{Bronski,Konotop}. Then, switching off the laser beams,
producing the trap, will result in a periodic distribution of the
condensate. However, it is not stable without the trap since it is not a solution
of the respective GP equation. This difficulty can be overcome if
simultaneously with switching off the optical trap one abruptly changes the
scattering length (or alternatively provides change of the number of particles)
in such a way
that the existing distribution will satisfy (\ref{NLS}). To be more specific,
let us consider an example of a BEC with a positive scattering length in an
optical trap given by \cite{Bronski}: $V(x)=-2V_0\sn(2\kappa x, m)$, where
$V_0$ is the potential amplitude, $\kappa=\lambda_1+\lambda_2$
and $m$ is the same as in (\ref{sn-dens}). The equation describing BEC
evolution now admits a solution
\begin{eqnarray*}
u(x,t)=\sqrt{V_0+(\la_1-\la_2)^2}\,e^{2i(\la_1^2+\la_2^2)t}\sn[(\la_1+\la_2)x,m].
\end{eqnarray*}
This last function solves also (\ref{NLS}) with $\sigma=1$ and
$
g=\frac{V_0+(\la_1-\la_2)^2}{(\la_1-\la_2)^2}
$
and thus by switching off the potential $V(x)$ with simultaneous changing
the scattering length by $\Delta a_s=\frac{a_0 V_0}{(\la_1-\la_2)^2}$
one achieves the desired initial state. Notice that although experimentally a sn-potential is not easily reachable in a general case, for a large range of  $m$ it is  approximated very well by only a few first Fourier harmonics. For example, for a situation described in Fig.~\ref{figone} one has $V(x)\approx -2V_0[1.47\sin(0.78 x)+0.15 \sin(2.3 x)]$ with the accuracy about 1\%.

In order to estimate characteristic scales of 
adiabatic deformations we introduce an
``aspect ratio" defined as $\delta=|\Delta
A|/L$, where $\Delta A$ is a total variation of the amplitude of
the periodic wave and $L$ is its wavelength. Then the cases $\delta\gg
1$ and $\delta \ll 1$ correspond to a solitonic lattice and to a
modulated plane wave. Dependence of the
scattering length on time can be simulated by $g(t)=e^{t/t_0}$ (physical units). Taking as an example the solution depicted in Fig.~\ref{figone}b, where $\delta\approx 1$ at $t=0$, we find that already at $t\approx 30$ms the
aspect ratio becomes $\delta\approx 100$. The
adiabaticity of the process means here that $\frac{\pi}{\omega_0}\gg t \gg \frac{\pi}{\omega_\bot}$ (physical units). It is satisfied well enough for traps with $\omega_\bot \gtrsim 2\pi\times 200$ Hz and $\omega_0 \lesssim 2\pi\times 5$ Hz. We notice that the imposed condition rules out a possibility of resonant phenomena which might be related to coincidence of a linear oscillator period, originated by one of the trap dimensions, on the one hand and the characteristic time of change of $a_s$ on the other hand. Stability of the above solutions has been studied numerically
in \cite{CKR}, where it has been found that soliton trains
are stable in the case of positive scattering length and are also stable
in the case of negative scattering length for special choice of the
parameters. In this context, the adiabatic variation
of the scattering length, which results in the change of the
wave parameters, can be used for stabilizing (or destabilizing)
respective periodic solutions.

To conclude, we have outlined the main idea of management of
periodic nonlinear waves in BEC's. The theory, although being developed
for a homogeneous NLS equation, give an accurate estimate of the central
part of a BEC placed in a magnetic trap, the later being studied numerically.
The existence of trap can also be taken into account in the framework of more
sophisticated theory recently developed in \cite{kamch02}.
Application of this method to the theory of a BEC will be presented elsewhere.

%\vskip 0.5cm
%A.M.K. is grateful to the staff of Centro de F\'{\i}sica da Mat\'eria
%Condensada, Universidade de Lisboa, for kind hospitality.
Work of A.M.K. in Lisbon has been supported by the Senior NATO Fellowship.
A.M.K. thanks also RFBR (Grant 01--01--00696) for partial support.
V.V.K. acknowledges support from the European grant, COSYC n.o.
HPRN-CT-2000-00158. Work of V.A.B. has been supported by the FCT fellowship SFRH/BPD/5632/2001. Cooperative work has been supported by the
NATO-Linkage grant No. PST.CLG.978177.

%\begin{references}


\begin{thebibliography}{99}
\bibitem{com} Our choice of the parameters is dictated by simplicity
of the obtained equations. Physical parameters can be expressed in terms of
$\lambda_i$ (see (\ref{cn}), (\ref{dn}) or (\ref{sn})).
\end{thebibliography}
\end{document}